\documentclass[journal]{IEEEtran}

\usepackage[noadjust]{cite}
\usepackage{graphicx}
\usepackage{float}
\usepackage{bm}
\usepackage{amsmath}
\usepackage{enumerate}
\usepackage{amssymb}
\usepackage{fixltx2e}
\usepackage{color}
\usepackage{multirow}

\title{Space Shift Keying with Reconfigurable Intelligent Surfaces: Phase Configuration Designs and Performance Analysis}

\author{Qiang~Li,~\IEEEmembership{Member,~IEEE}, Miaowen Wen,~\IEEEmembership{Senior Member,~IEEE}, Shuai Wang,~\IEEEmembership{Member,~IEEE}, \\
George C. Alexandropoulos,~\IEEEmembership{Senior Member,~IEEE}, and Yik-Chung Wu,~\IEEEmembership{Senior Member,~IEEE}
\thanks{The work was supported in part by the National Natural Science Foundation of China under Grant 61871190 and Grant 62001203, in part by the Natural Science Foundation of Guangdong Province under Grant 2018B030306005 and Grant 2020A1515110470, in part by the Pearl River Nova Program of Guangzhou under Grant 201806010171, and in part by the Fundamental Research Funds for the Central Universities under Grant 2019SJ02. \textit{(Corresponding author: Miaowen Wen.)}}
\thanks{Qiang Li is with the Department of Electronic Engineering, College of Information Science and Technology, Jinan University, Guangzhou 510632, China (e-mail: qiangli@jnu.edu.cn).}
\thanks{Miaowen Wen is with the School of Electronic and Information Engineering,
South China University of Technology, Guangzhou 510640, China (e-mail: eemwwen@scut.edu.cn).}
\thanks{Shuai Wang is with the Department of Electrical and Electronic Engineering, and the Department of Computer
Science and Engineering, Southern University of Science and Technology, Shenzhen 518055, China, and also with the Sifakis Research Institute
of Trustworthy Autonomous Systems, Shenzhen 518055, China (e-mail: wangs3@sustech.edu.cn).}
\thanks{George C. Alexandropoulos is with the Department of Informatics and Telecommunications, National and Kapodistrian University of Athens, Panepistimiopolis Ilissia, 15784 Athens, Greece (e-mail: alexandg@di.uoa.gr).}
\thanks{Yik-Chung Wu is with the Department of Electrical and Electronic Engineering, The University of Hong Kong, Hong Kong (e-mail:
ycwu@eee.hku.hk).}
}

\begin{document}
\maketitle

%\IEEEtitleabstractindextext{
\begin{abstract}
Reconfigurable intelligent surface (RIS)-assisted transmission and space shift keying (SSK) appear as promising candidates for future energy-efficient wireless systems. In this paper, two RIS-based SSK schemes are proposed to efficiently improve the error and throughput performance of conventional SSK systems, respectively. The first one, termed RIS-SSK with passive beamforming (RIS-SSK-PB), employs an RIS for beamforming and targets the maximization of the minimum squared Euclidean distance between any two decision points. The second one, termed RIS-SSK with Alamouti space-time block coding (RIS-SSK-ASTBC), employs an RIS for ASTBC and enables the RIS to transmit its own Alamouti-coded information while reflecting the incident SSK signals to the destination. A low-complexity beamformer and an efficient maximum-likelihood (ML) detector are designed for RIS-SSK-PB and RIS-SSK-ASTBC, respectively. Approximate expressions for the average bit error probabilities of the source and/or the RIS are derived in closed-form assuming ML detection. Extensive computer simulations are conducted to verify the performance analysis. Results show that RIS-SSK-PB significantly outperforms the existing RIS-free and RIS-based SSK schemes, and RIS-SSK-ASTBC enables highly reliable transmission with throughput improvement.
\end{abstract}

%}

\section{Introduction}
\PARstart{T}HE deployment of the fifth generation (5G) cellular networks is accelerating across the world. This wireless communication standard is expected to support lots of new applications and services, which require various enabling technologies. Among these, energy-efficient transmission is a key enabler for energy-constrained networks, such as Internet of Things. However, the traditional digital modulation schemes that alter the amplitude, phase, and/or the frequency of a sinusoidal carrier signal often involve complicated operations, such as mixing and filtering.

Recently, the concept of index modulation (IM) that leverages upon the ON/OFF state of the transmission entities to convey information has created completely new dimensions for energy-efficient transmission \cite{Wen2017Index,Cheng2018Index,Basar2017Index,Li2020Subcarrier}. As a prominent member of IM, spatial modulation (SM) uses the transmit antennas of a multiple-input multiple-output (MIMO) system in an innovative fashion \cite{Wen2019A,Renzo2014Spatial,Mesleh2008Spatial}. In SM, only a single transmit antenna is activated for each transmission of a constellation symbol, and the index of the active antenna is used to convey extra information bits. By limiting the active antenna to simply transmit an unmodulated sinusoidal carrier signal, SM evolves into space shift keying (SSK) that employs the index of the active antenna only for information transfer \cite{Jeganathan2009Space}. Since the traditional signal modulation is avoided, SSK achieves high energy efficiency and enjoys low transceiver complexity. Therefore, SSK is a promising candidate for future energy-efficient wireless communications. However, SSK still faces challenges. On one hand, SSK-based solutions are based on the fact that different active antennas lead to different channel realizations, whereas they cannot configure the wireless propagation environment itself. The system performance of SSK highly depends on the distinctness of the channel signatures associated with different active antennas. Hence, rich scattering in the propagation environment and/or a large number of receive antennas are required for SSK to avoid poor error performance. Unfortunately, these requirements may not be satisfied in general. On the other hand, the sinusoidal carrier signal itself transmitted from the active antenna does not carry any information, resulting in low system throughput. For SSK, embedding information in the carrier signal while avoiding traditional digital modulation is still an open issue.

The emerging technology of reconfigurable intelligent surfaces (RISs) has also received a lot of research interests due to the distinctive capability of turning an uncontrollable and unfavorable environment into a controllable and benign entity \cite{Renzo2020Smart,Renzo2019Smart,Basar2019Wireless}. An RIS is a man-made planar surface consisting of a large number of reflecting elements. Each of those elements is able to reflect the incident signals by applying an adjustable phase shift. This dynamically adjustable reflection, that goes beyond Snell's law, is accomplished without any form of the conventional power amplifiers, since no new signal is generated at the RIS side, and with no contamination of the impinging signal with reception thermal noise \cite{Huang2020Holographic}. In the literature, an RIS is often deployed to modify channel links and/or to transmit its own information \cite{Tang2020Wireless}. Specifically, to modify channel links, the RIS-based dual-hop scheme was proposed in \cite{Basar2019Transmission} for single-input single-output systems, where the RIS is adjusted to maximize the instantaneous receive signal-to-noise ratio (SNR). In \cite{Ye2020Joint}, an RIS was utilized to minimize the symbol error rate of precoded spatial multiplexing MIMO systems via the joint reflecting and precoding optimization. A cosine similarity-based low-complexity reflecting optimization was proposed in \cite{Yigit2020Low} for RIS-assisted spatial multiplexing MIMO systems by maximizing the overall channel gain. The authors in \cite{Dong2020Enhancing} employed an RIS to enhance the secrecy rate of MIMO transmission in the presence of an eavesdropper by jointly optimizing the transmit covariance and reflection coefficient matrices. References \cite{Wu2019Intelligent,Wu2020Beamforming,Huang2019Reconfigurable,Huang2018Energy} applied the RIS technology to multiuser downlink multiple-input single-output systems. In \cite{Wu2019Intelligent} and \cite{Wu2020Beamforming}, the transmit power was minimized via the joint optimization of the active beamforming at the base station and the passive beamforming (PB) at the RIS, by allowing continuous and discrete reflection coefficients, respectively. Aiming to maximize the energy efficiency, the transmit power for each user and the reflection coefficients with both infinite \cite{Huang2019Reconfigurable} and realistic low-resolution \cite{Huang2018Energy} phase configuration codebooks were jointly optimized. Due to the attractive advantages, researchers are also utilizing RISs to enhance other existing techniques, such as non-orthogonal multiple access \cite{Hou2020Reconfigurable} and orthogonal frequency division multiplexing \cite{Zheng2020Intelligent}. As the authors in \cite{Bouida2020Reconfigurable} applied reconfigurable antennas to SSK, the authors in \cite{Canbilen2020Reconfigurable} proposed the intelligent RIS-SSK, which makes real-time adjustments to the reflection coefficients for maximizing the instantaneous receive SNR provided that the active antenna index is known to the RIS per transmission.

Besides solely acting as a signal reflector, an RIS can be employed for information transfer by adjusting the reflection coefficients \cite{Li2020Single}. For an RIS-aided channel, the capacity-achieving scheme was demonstrated to jointly encode information in the RIS configuration as well as in the transmitted signal \cite{Karasik2020Beyond}. Both phase shift keying (PSK) and quadrature amplitude modulation can be achieved at each reflecting element, thus providing a new design of MIMO transmission \cite{Tang2020MIMO}. In \cite{Guo2020Reflecting}, the reflecting modulation that uses the indices of reflecting patterns for information embedding was introduced into traditional MIMO frameworks. In \cite{Yan2020Passive} and \cite{Lin2020Reconfigurable}, the SM principle was applied to an RIS, i.e., the information is carried by the ON/OFF status of reflecting elements. In \cite{Basar2020Reconfigurable}, the PB was performed at the RIS to steer the signal towards a certain receive antenna, thus using the index of the selected receive antenna to convey information. Based on \cite{Basar2020Reconfigurable}, \cite{Ma2020Large} further applied SM to the RF side, leading to even higher system throughput. Similar to the antenna-based MIMO, Alamouti space-time block coding (ASTBC) and Vertical Bell Labs layered space-time (VBLAST) schemes was implemented with an RIS in \cite{Khaleel2020Reconfigurable}.

Motivated by the challenges faced by SSK and the recently identified potential of the amalgamation of RISs and SSK, we study RIS-based SSK in this paper towards future energy-efficient wireless communications. Two RIS-based SSK schemes are proposed to improve the error and throughput performance of conventional SSK systems, respectively. The contributions of this paper are summarized as follows.
\begin{itemize}
    \item The first proposed scheme, which is called RIS-SSK with PB (RIS-SSK-PB), employs an RIS for PB and targets the maximization of the minimum squared Euclidean distance between any two decision points. Compared with the intelligent RIS-SSK scheme in \cite{Canbilen2020Reconfigurable}, RIS-SSK-PB achieves the following advantages: 1) the knowledge of the active antenna index is not required any more at the RIS; 2) the reflection coefficients are not adjusted online per transmission any more; and 3) higher diversity gains can be obtained with the same detection complexity. We devise a semi-definite relaxation (SDR) method for the considered RIS reflection optimization problem as well as a low-complexity algorithm. We analyze the bit error rate (BER) performance of RIS-SSK-PB over Rayleigh fading channels assuming two transmit antennas and maximum-likelihood (ML) detection. An approximate expression for the average bit error probability (ABEP) of RIS-SSK-PB is also derived in closed-form.

    \item The second proposed scheme, which is called RIS-SSK with ASTBC (RIS-SSK-ASTBC), employs an RIS for ASTBC and enables the RIS to transmit its own Alamouti-coded information, while reflecting the incident SSK signals to the destination. We design a low-complexity ML detector for RIS-SSK-ASTBC. The BER performance of RIS-SSK-ASTBC with ML detection over Rayleigh fading channels is analyzed. Approximate and asymptotic ABEP expressions are derived in closed-form for the source and the RIS, which reveal that their information bits have two-diversity-order protection with increasing SNR or the number of reflecting elements.

\end{itemize}

The rest of this paper is organized as follows. The transceiver structure and the performance analysis of RIS-SSK-PB are presented in Section II. Section III describes the scheme of RIS-SSK-ASTBC, including the system model and performance analysis. Section IV presents computer simulation results, followed by the conclusion in Section V.

\textit{Notation:} Column vectors and matrices are in the form of lowercase and capital bold letters, respectively. Superscripts $^*$, $^T$, and $^H$ stand for conjugate, transpose, and Hermitian transpose, respectively. $j=\sqrt{-1}$ denotes the imaginary unit. $\mathrm{tr}(\cdot)$ and $\mathrm{rank}(\cdot)$ return the trace and rank of a matrix, respectively. $\mathrm{diag}(\cdot)$ transforms a vector into a diagonal matrix. $(\mathcal{C})\mathcal{N}(\mu,\sigma^2)$ represents the (complex) Gaussian distribution with mean $\mu$ and variance $\sigma^2$. The probability of an event and the probability density function (PDF) of a random variable are denoted by $\Pr(\cdot)$ and $f(\cdot)$, respectively. $Q(\cdot)$ represents the Gaussian $Q$-function. $E\{\cdot\}$ and $Var\{\cdot\}$ denote expectation and variance, respectively. $|\cdot|$, $\angle$, and $\Re\{\cdot\}$ denote the absolute, phase, and real part of a complex number, respectively. $f_i$ denotes the $i$-th entry of $\mathbf{f}$, $\mathbf{g}_l$ represents the $l$-th column of $\mathbf{G}$, and $g_{il}$ stands for the $(i,l)$-th element of $\mathbf{G}$. $\|\cdot\|$ denotes the Frobenius norm. $\mathbf{V} \succeq 0$ means that $\mathbf{V}$ is positive semi-definite.

\section{The RIS-SSK-PB Scheme}
\begin{figure}[t]
	\centering
	\includegraphics[width=3.4in]{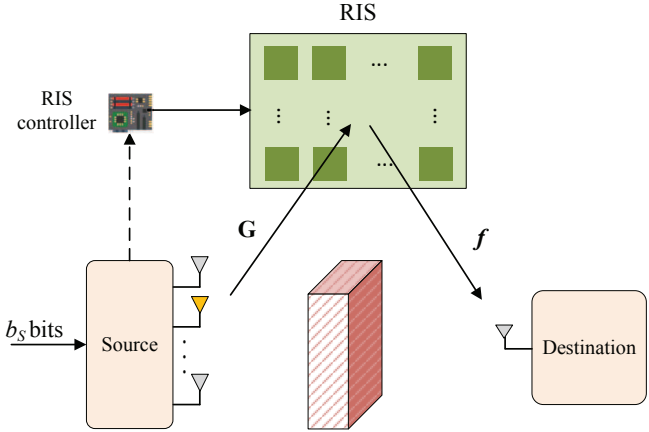}
	\caption{System model of RIS-SSK-PB.}
	\label{Fig1}
\end{figure}
In this section, we study the proposed RIS-SSK-PB scheme, in which the RIS merely reflects the incident SSK signals without transmitting its own information.

\subsection{System Model}
Fig.~\ref{Fig1} depicts the system model of RIS-SSK-PB, which consists of a source, a destination, and an RIS. Due to an obstacle, the source and the destination communicate with each other through the RIS. The source is equipped with $N_t$ transmit antennas, and the destination adopts single-antenna reception for a simple receiver structure. The RIS is connected to a controller and made up of $N$ reflecting elements that are deployed on a two-dimensional rectangular grid. The channel matrices of the source-to-RIS and RIS-to-destination links are denoted by $\mathbf{G} \in \mathbb{C}^{N \times N_t}$ and $\mathbf{f} \in \mathbb{C}^{N \times 1}$, respectively. Specifically, $g_{il}$ represents the channel coefficient between the $l$-th transmit antenna and the $i$-th reflecting element, while $f_i$ is the channel coefficient between the $i$-th reflecting element and the destination, where $i \in \{1,\ldots,N\}$ and $l \in \{1,\ldots,N_t\}$. The reflecting elements are assumed placed with horizontal and vertical inter-element distances equal or greater than half the signal wavelength. In this special case, the fading channels can be approximately considered as independent and identically distributed (i.i.d.) fading, i.e., $g_{il}$ and $f_i$ are i.i.d. complex Gaussian random variables following the distribution $\mathcal{CN}(0,1)$ for $i=1,\ldots,N$ and $l=1,\ldots,N_t$ \cite{Huang2019Reconfigurable,Canbilen2020Reconfigurable}. We note that the practical case of correlated fading \cite{Bjornson2020Rayleigh} will be considered in future work. The reflection coefficient for the $i$-th reflecting element of the RIS can be expressed as $\phi_i=\beta_i\exp(j\theta_i)$, where $\beta_i \in [0,1]$ and $\theta_i \in [0,2\pi)$ are the amplitude coefficient and phase shift, respectively. To characterize the performance limit of RIS-SSK-PB, we assume that $\beta_i=1$, $\theta_i$ can be continuously varied in $[0,2\pi)$ for $i=1,\ldots,N$, and the channel state information (CSI) of the source-to-RIS and RIS-to-destination links is perfectly known to the RIS and the destination. Note that the channel estimation in RIS-empowered wireless communications is a challenging problem, however, with encouraging approaches already \cite{Zheng2020Intelligent,Alexandropoulos2020A}.

For each transmission, the source encodes $b_S$ bits into an index $l\in\{1,\ldots,N_t\}$, and then emits an unmodulated carrier signal from the antenna with the index $l$ towards the RIS, resulting in the SSK signal
\begin{align}\label{SSK_x}
{\mathbf{x}} = \big[\underbrace {0, \ldots ,0,}_{l - 1}1,\underbrace {0, \ldots ,0}_{{N_t} - l}\big]^T,
\end{align}
where $b_S=\log_2(N_t)$, and $N_t$ is assumed to be an arbitrary integer power of two, such that $l$ can be easily obtained by converting the $b_S$ bits into the decimal representation. The RIS reflects the incident signal with reflection vector $\bm{\phi}=[\exp(j\theta_1),\ldots,\exp(j\theta_N)]^T \in \mathbb{C}^{N \times 1}$, leading to the received baseband signal at the destination as follows:
\begin{align}\label{Scheme1_y1}
y = \mathbf{f}^T\mathbf{\Phi Gx} + w=\mathbf{f}^T\mathbf{\Phi g}_l + w,
\end{align}
where $\mathbf{\Phi}=\mathrm{diag}(\bm{\phi})$ and $w$ is the additive white Gaussian noise (AWGN) at the destination, which follows the distribution $\mathcal{CN}(0,N_0)$. We define the transmit signal-to-noise ratio (SNR) as $\rho = 1/N_0$. With (\ref{Scheme1_y1}), the destination adopts ML detection to decode $l$, namely
\begin{align}\label{Scheme1_ML}
\hat l = \arg \mathop {\min }\limits_l{\left| {y - \mathbf{f}^T\mathbf{\Phi g}_l} \right|^2},
\end{align}
where $\hat l$ is the estimate of $ l$. The destination then recovers $b_S$ information bits from $\hat l$. We observe from (\ref{Scheme1_ML}) that the value of $\bm{\phi}$ undoubtedly influences the error performance. In the next subsection, we will focus on the optimization of $\bm{\phi}$.

\subsection{Reflecting Optimization}
In the intelligent RIS-SSK scheme of \cite{Canbilen2020Reconfigurable}, $\bm{\phi}$ is optimized by maximizing the instantaneous receive SNR at the destination. More specifically,
the received signal in (\ref{Scheme1_y1}) can be rewritten as
\begin{align}\label{Scheme1_y2}
y = \sum\limits_{i = 1}^N {{f_i}{g_{il}}\exp \left( {j{\theta _i}} \right)} + w.
\end{align}
Obviously, the instantaneous receive SNR is given by $\rho| \sum_{i=1}^{N}f_ig_{il}\exp(j\theta_i)|^2$, which is maximized when $\theta _i=- \angle {f_i} - \angle {g_{il}}$. This optimization criterion, however, requires that the knowledge of the active antenna index be available at the RIS in real time, and that $\bm{\phi}$ be adjusted online per transmission even though the CSI remains unchanged, which is unrealistic. Moreover, since the BER performance of SSK highly depends on the channel difference associated with different active transmit antennas \cite{Wen2019A,Renzo2014Spatial}, maximizing the receive SNR may not lead to good performance. These observations motivate us to propose other optimization algorithms.

Let us first examine the conditional pairwise error probability (PEP), which is defined as the probability of detecting $l$ incorrectly as $\hat{l}$ conditioned on $\mathbf{G}$, $\mathbf{f}$, and $\bm{\phi}$, namely
\begin{align}
\Pr \left( {l \to \hat l\left| {{\bf{G}},{\bf{f}},\bm{\phi} } \right.} \right) &= \Pr \left( {{{\left| {y - {{\bf{f}}^T}{\bf{\Phi }}{{\bf{g}}_l}} \right|}^2} > {{\left| {y - {{\bf{f}}^T}{\bf{\Phi }}{{\bf{g}}_{\hat l}}} \right|}^2}} \right) \nonumber \\
& = \Pr \Big( -{{\left| {{{\bf{f}}^T}{\bf{\Phi }}\left( {{{\bf{g}}_l} - {{\bf{g}}_{\hat l}}} \right)} \right|}^2} \nonumber \\
&\;\; -2\Re\left\{w^*\left[\mathbf{f}^T\mathbf{\Phi}(\mathbf{g}_l - \mathbf{g}_{\hat l})\right]\right\} > 0 \Big) \nonumber \\
& = \Pr \left(D>0\right),
\end{align}
where $D$ is a Gaussian random variable with
\begin{align}
E\{D\}&= -{{\left| {{{\bf{f}}^T}{\bf{\Phi }}\left( {{{\bf{g}}_l} - {{\bf{g}}_{\hat l}}} \right)} \right|}^2}, \\
Var\{D\}&= 2N_0{{\left| {{{\bf{f}}^T}{\bf{\Phi }}\left( {{{\bf{g}}_l} - {{\bf{g}}_{\hat l}}} \right)} \right|}^2}.
\end{align}
Hence, we have
\begin{align}\label{CPEP_PB}
\Pr \left( {l \to \hat l\left| {{\bf{G}},{\bf{f}},\bm{\phi} } \right.} \right)=Q\left( {\sqrt {\frac{{\rho {{\left| {{{\bf{f}}^T}{\bf{\Phi }}\left( {{{\bf{g}}_l} - {{\bf{g}}_{\hat l}}} \right)} \right|}^2}}}{2}} } \right).
\end{align}
Then, according to the union bounding technique \cite{Proakis2005Digital}, an upper bound on the instantaneous error probability of active antenna index detection can be expressed as
\begin{align}\label{Scheme1_UB}
{P_e} &\le \frac{2}{{{N_t}}}\sum\limits_{l = 1}^{{N_t}} {\sum\limits_{\scriptstyle\hat l = 1\hfill\atop
\scriptstyle l < \hat l\hfill}^{{N_t}} \Pr \left( {l \to \hat l\left| {{\bf{G}},{\bf{f}},\bm{\phi} } \right.} \right) } \nonumber \\
&\le (N_t-1) \cdot Q\left( {\sqrt {\frac{{{\rho}}}{{2}}d_{\min }} } \right),
\end{align}
where $d_{\min }$ is defined as
\begin{align}\label{Scheme1_dmin}
d_{\min } = \mathop {\min }\limits_{\scriptstyle l,\hat l\hfill\atop
\scriptstyle l \ne \hat l\hfill} {\left| {{{\bf{f}}^T}{\bf{\Phi }}\left( {{{\bf{g}}_l} - {{\bf{g}}_{\hat l}}} \right)} \right|^2}.
\end{align}
Therefore, from (\ref{Scheme1_UB}), we propose to optimize $\bm{\phi}$ by maximizing $d_{\min }$, namely
\begin{align}\label{Scheme1_P1}
\mathrm{(P1)}: \quad & \mathop {\max }\limits_{\bm{\phi}} \; {d_{\min }} \\
& s.t. \; \theta_i \in [0,2\pi), \quad i=1,\ldots,N.
\end{align}
For $N_t=2$, the optimal solution to problem (P1) can be easily derived in closed-form as
\begin{align}\label{phi_opt_2}
{\bm{\phi}}_{opt} &= \arg \mathop {\max }\limits_{\bm{\phi}} {\left| {{{\bf{f}}^T}{\bf{\Phi }}\left( {{{\bf{g}}_1} - {{\bf{g}}_2}} \right)} \right|^2} \nonumber \\
&= \arg \mathop {\max }\limits_{\bm{\phi}} {\left| {\sum\limits_{i = 1}^N {{f_i}} \left( {{g_{i1}} - {g_{i2}}} \right)\exp \left( {j{\theta _i}} \right)} \right|^2} \nonumber \\
&= [-\angle {f_1} - \angle {(g_{11}-g_{12})}, \nonumber \\
&\quad \quad \ldots, -\angle {f_N} - \angle {(g_{N1}-g_{N2})}]^T.
\end{align}
Unfortunately, for $N_t > 2$, it is not an easy task to solve problem (P1) optimally due to the max-min criterion of (\ref{Scheme1_P1}). Therefore, we first apply the SDR method and then propose a low-complexity algorithm to solve problem (P1) sub-optimally in the following.

\subsubsection{SDR Method}
For SDR, by introducing an auxiliary variable $t$, problem (P1) can be equivalently reformulated as
\begin{align}
\mathrm{(P2)}: \quad &\mathop {\max }\limits_{\bm{\phi}} \; {t} \\
&s.t. \; {\left| {{{\bf{f}}^T}{\bf{\Phi }}\left( {{{\bf{g}}_l} - {{\bf{g}}_{\hat l}}} \right)} \right|^2} \ge t, \quad \forall l,\hat{l}=1,\ldots,N_t, l \ne \hat{l},\label{Scheme1_P2_C1}\\
&\theta_i \in [0,2\pi), \quad \forall i=1,\ldots,N.
\end{align}
However, problem (P2) is still non-convex due to the constraint in (\ref{Scheme1_P2_C1}). Actually, the term on the left-hand side of (\ref{Scheme1_P2_C1}) can be expressed as
\begin{align}
{\left| {{{\bf{f}}^T}{\bf{\Phi }}\left( {{{\bf{g}}_l} - {{\bf{g}}_{\hat l}}} \right)} \right|^2} &= {\left| {{{\bf{f}}^T}\mathrm{diag}\left( {{{\bf{g}}_l} - {{\bf{g}}_{\hat l}}} \right){\bm{\phi}}} \right|^2} \nonumber \\
&= {{\bm{\phi}}^H}\mathbf{R}{\bm{\phi}} \nonumber \\
&= \mathrm{tr}\left( {\mathbf{R}{\bm{\phi}}{{\bm{\phi}}^H}} \right),
\end{align}
where $\mathbf{R} = {({{\bf{f}}^T}\mathrm{diag}( {{{\bf{g}}_l} - {{\bf{g}}_{\hat l}}} ))^H}({{\bf{f}}^T}\mathrm{diag}( {{{\bf{g}}_l} - {{\bf{g}}_{\hat l}}}))$. Define $\mathbf{V} =\bm{\phi}\bm{\phi}^H$ with $\mathbf{V} \succeq 0 $ and $\mathrm{rank}(\mathbf{V}) = 1$. Since the rank-one constraint is non-convex, we use SDR to relax this constraint. As a result, problem (P2) is reduced to
\begin{align}\label{Scheme1_P2}
\mathrm{(P3)}: \quad &\mathop {\max }\limits_{\bm{V}} \; {t} \\
& s.t. \; \mathrm{tr}\left( \mathbf{RV} \right)\ge t, \quad \forall l,\hat{l}=1,\ldots,N_t, l \ne \hat{l},\\
&V_{ii}=1, \quad \forall i=1,\ldots,N,\\
&\mathbf{V} \succeq 0.
\end{align}

Obviously, problem (P3) is a convex semi-definite program and can be solved by existing convex optimization solvers, such as CVX \cite{Grant2016Intelligent}. However, the rank-one constraint is relaxed in problem (P3), such that the solution may not satisfy $\mathrm{rank}(\mathbf{V}) = 1$. This implies that the optimal objective value of problem (P3) is an upper bound on that of problem (P1). Therefore, extra steps are required to obtain a rank-one solution from the solution to problem (P3).
Specifically, after taking the eigenvalue decomposition of $\mathbf{V}$ as $\mathbf{V} = \mathbf{U}\mathbf{\Sigma} \mathbf{U}^H$, where $\mathbf{U} \in \mathbb{C}^{N \times N}$ and $\mathbf{\Sigma} \in \mathbb{C}^{N \times N}$ are a unitary matrix and a diagonal matrix, respectively, we can obtain a sub-optimal solution to problem (P3) as $\bm{\phi}= \mathbf{U}\mathbf{\Sigma}^{1/2}\mathbf{r}$, where $\mathbf{r} \in \mathbb{C}^{N \times 1}$ is a random vector with each element drawn from $\mathcal{CN}(0,1)$. Further, by generating a sufficiently large number of realizations of $\mathbf{r}$ and selecting the one, denoted by $\tilde{\mathbf{r}}$, that maximizes $d_{\min}$, the sub-optimal solution to problem (P1) can be derived as
$\bm{\phi}=[\tilde{\phi}_1/|\tilde{\phi}_1|,\ldots,\tilde{\phi}_N/|\tilde{\phi}_N|]^T$, where $[\tilde{\phi}_1,\ldots,\tilde{\phi}_N]^T=\mathbf{U}\mathbf{\Sigma}^{1/2}\tilde{\mathbf{r}}$.

\subsubsection{Low-Complexity Algorithm}
To achieve lower complexity than the SDR method, we propose a low-complexity algorithm here. The core idea is to derive a set of $\bm{\phi}$, denoted by $\mathbb{A}$, from which the $\bm{\phi}$ capable of maximize $d_{\min}$ in (\ref{Scheme1_dmin}) is selected as the solution to problem (P1). Obviously, $\mathbb{A}$ is the key. Inspired by the solution to the case of $N_t=2$, a heuristic $\mathbb{A}$ with $N_t(N_t-1)/2$ entities can be constructed by $\bm{\phi}_{ij}=[-\angle {f_1} - \angle {(g_{1i}-g_{1j})}, \ldots, -\angle {f_N} - \angle {(g_{Ni}-g_{Nj})}]^T$, where $i,j \in \{1,\ldots,N_t\},i<j$. Note that since the optimal $\bm{\phi}$ may not be included in $\mathbb{A}$, the low-complexity algorithm is suboptimal. However, as will be shown in Section~V, compared with the SDR method, this algorithm can achieve satisfying performance with much reduced complexity.

\subsection{Performance Analysis for $N_t=2$}
Since the reflection matrix $\bm{\Phi}$ is directly correlated with the channel matrices $\mathbf{f}$ and $\mathbf{G}$, as seen from Section II.B,
it is very difficult to derive the distribution of the composite channel matrix $\mathbf{f}^T\mathbf{\Phi G}$ and analyze the BER performance of RIS-SSK-PB for a general case of $N_t$. Therefore, we focus only on the performance analysis for $N_t=2$ to obtain insights.

For $N_t=2$, the optimal $\bm{\phi}$ is given in (\ref{phi_opt_2}). From (\ref{CPEP_PB}), the instantaneous BEP of RIS-SSK-PB can be expressed as
\begin{align}
{P_{bi}} = Q\left( {\sqrt {\frac{{\rho {{\left| v \right|}^2}}}{2}} } \right),
\end{align}
where $v = \sum\nolimits_{i = 1}^N {\left| {{f_i}} \right|\left| {{g_{i1}} - {g_{i2}}} \right|} $. Under the assumption of $N \gg 1$, and according to the central limit theorem (CLT) \cite{Boulogeorgos2020Performance}, $v$ can be regarded as a Gaussian random variable following the distribution $\mathcal{N}(\mu_v,\sigma_v^2)$, where $\mu_v$ and $\sigma_v^2$ can be derived as follows:
\begin{align}
\mu_v=\frac{{\sqrt 2 }}{4} \pi N,\quad \sigma_v^2=\left(2-\frac{\pi^2}{8} \right)N.
\end{align}
Hence, $|v|^2$ is a noncentral chi-square random variable with one degree of freedom and noncentrality parameter $a^2=\pi^2N^2/8$, whose moment generating function is given by
\begin{align}
{\Psi _{|v|^2}}\left( s \right) = {\left( {\frac{1}{{1 - 2\sigma _v^2s}}} \right)^{1/2}}\exp \left( {\frac{{{a^2}s}}{{1 - 2\sigma _v^2s}}} \right).
\end{align}
Then, the ABEP of RIS-SSK-PB with $N_t=2$ can be obtained by averaging ${P_{bi}}$ over $|v|^2$, namely
\begin{align}
{P_b} &= {E_{|v{|^2}}}\left\{ {{P_{bi}}} \right\} \approx \frac{1}{{12}}{\Psi _{|v|^2}}\left( { - \frac{\rho }{4}} \right) + \frac{1}{4}{\Psi _{|v|^2}}\left( { - \frac{\rho }{3}} \right), \nonumber \\
&= \frac{1}{12}{\left( {\frac{2}{{2 + \sigma _v^2\rho}}} \right)^{1/2}}\exp \left(-{\frac{{{a^2}\rho}}{{4 + 2\sigma _v^2\rho}}} \right) \nonumber \\
&\hspace{+0.4cm}+\frac{1}{4}{\left( {\frac{3}{{3 + 2\sigma _v^2\rho}}} \right)^{1/2}}\exp \left(-{\frac{{{a^2}\rho}}{{3 + 2\sigma _v^2\rho}}} \right),
\end{align}
where the approximation results from \cite{Chiani2003New}
\begin{align}\label{Q_fun}
Q\left( x \right) \approx \frac{1}{12} \exp\left(- \frac{x^2}{2}\right) + \frac{1}{4} \exp\left(- \frac{2x^2}{3}\right).
\end{align}
Since $\sigma _v^2$ and $a^2$ grows linearly and quadratically with increasing $N$, respectively, we observe from $P_b$ that the performance improvement achieved by doubling $N$ is greater than that by doubling $\rho$, i.e., doubling $N$ results in SNR gains greater than 3 dB. Moreover, as $\exp(-x)$ is a concave function, the SNR gains achieved by doubling $N$ are increasingly small.

\textit{Remark 1:} Since the proposed reflecting optimization is aimed to maximize the minimum squared Euclidean distance between any two decision points, RIS-SSK-PB is expected to outperform the intelligent RIS-SSK in terms of BER. Moreover, the instantaneous knowledge of the active antenna index is not required at the RIS, and the reflection coefficients are not adjusted online per transmission, but only when any of the channel matrices change,  for RIS-SSK-PB. The drawback of RIS-SSK-PB is that the reflecting optimization involves high computational complexity in comparison with the intelligent RIS-SSK. Fortunately, the operation can be implemented offline, and the results remain unchanged until the CSI varies. In addition, the detection complexity and the requirement of CSI of RIS-SSK-PB are the same as those of the intelligent RIS-SSK.

\section{The RIS-SSK-ASTBC Scheme}
\begin{figure}[t]
	\centering
	\includegraphics[width=3.5in]{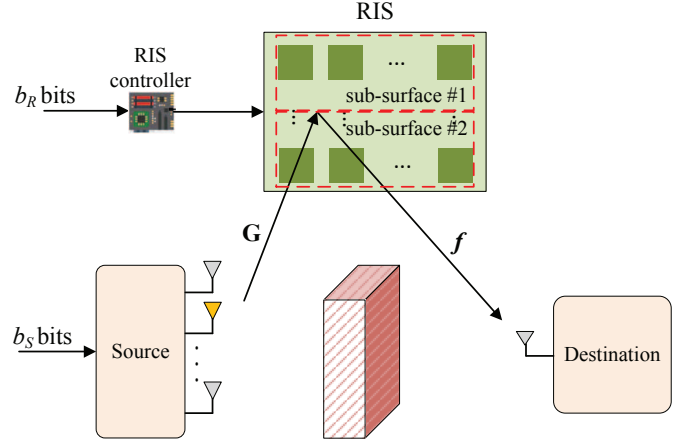}
	\caption{System model of RIS-SSK-ASTBC.}
	\label{Fig2}
\end{figure}
In this section, we present the RIS-SSK-ASTBC scheme, in which the RIS not only reflects the incident SSK signals but also transmits its own information via ASTBC.

\subsection{System Model}
The system model for RIS-SSK-ASTBC is depicted in Fig.~\ref{Fig2}, where the RIS with an even number of reflecting elements is divided into two sub-surfaces, namely sub-surface \#1 and sub-surface \#2, each containing $N/2$ reflecting elements. In RIS-SSK-ASTBC, the source transmits SSK symbols to the destination via the RIS symbol-by-symbol with guard intervals in between, while the RIS conveys its own information to the destination through the Alamouti transmission scheme. We assume that channel links experience small-scale flat Rayleigh fading as in RIS-SSK-PB. It should be noted that, different from the case in RIS-SSK-PB, the CSI of both links is required at the destination only, not at the RIS any more.

Each complete transmission is comprised of two time slots. In the first time slot, the RIS's own information of $b_R=2\log_2 (M)$ bits is loaded into the RIS controller. Specifically, the first $\log_2 (M)$ bits are used to adjust the phase shifts of the sub-surface \#1 to $\theta_i=\alpha_1$ with $i=1,\ldots,N/2$, and the remaining $\log_2 (M)$ bits are used to adjust the phase shifts of the sub-surface \#2 to $\theta_i=\alpha_2$ with $i=N/2+1,\ldots,N$, where $\alpha_1,\alpha_2 \in \mathcal{M}(=\{0,2\pi/M,\ldots,2\pi(M-1)/M\})$. In this manner, two virtual $M$-PSK symbols are generated at the RIS. Meanwhile, the source transmits $b_S=\log_2(N_t)$ bits via an SSK signal towards the RIS. Therefore, the received signal at the destination in the first time slot can be expressed as
\begin{align}\label{y1}
y_1= \exp (j\alpha_1)\sum\limits_{i = 1}^{N/2} {{f_i}{g_{il}}}  + \exp (j\alpha_2)\sum\limits_{i = N/2 + 1}^N {{f_i}{g_{il}}} +w_1,
\end{align}
where $l$ denotes the index of the active transmit antenna and $w_1$ represents the AWGN at the destination in the first time slot. The first and second summations of terms are from sub-surface \#1 and sub-surface \#2, respectively. In the second time slot, the phase shifts of the sub-surface \#1 are adjusted to $\theta_i=\pi-\alpha_2$ with $i=1,\ldots,N/2$, and the phase shifts of the sub-surface \#2 are adjusted to $\theta_i=-\alpha_1$ with $i=N/2+1,\ldots,N$. The source transmits the same SSK signal as that in the first time slot. Therefore, the received signal at the destination in the second time slot can be given by
\begin{align}\label{y2}
y_2= -\exp (-j\alpha_2)\sum\limits_{i = 1}^{N/2} {{f_i}{g_{il}}}  + \exp (-j\alpha_1)\sum\limits_{i = N/2 + 1}^N {{f_i}{g_{il}}} +w_2,
\end{align}
where $w_2$ denotes the AWGN at the destination in the second time slot. The spectral efficiency of RIS-SSK-ASTBC is $\log_2 (M) +\log_2(N_t)/2$ bits per channel use (bpcu), while that of traditional SSK is $\log_2(N_t)$ bpcu. Obviously, RIS-SSK-ASTBC achieves a higher spectral efficiency than traditional SSK in the case of $M > \sqrt{N_t}$.

After re-organizing (\ref{y1}) and (\ref{y2}) into a matrix form, we have
\begin{align}\label{y}
{\bf{y}} &= \left[ {{y_1}},{{y_2}} \right]^T \nonumber \\
&= \underbrace {\left[ {\begin{array}{*{20}{c}}
{\exp (j{\alpha _1})}&{\exp (j{\alpha _2})}\\
{ - \exp ( - j{\alpha _2})}&{\exp ( - j{\alpha _1})}
\end{array}} \right]}_{\bf{C}}\underbrace {\left[ {\begin{array}{*{20}{c}}
{\sum\limits_{i = 1}^{N/2} {{f_i}{g_{il}}} }\\
{\sum\limits_{i = N/2 + 1}^N {{f_i}{g_{il}}} }
\end{array}} \right]}_{{{\bf{h}}_l}} + {\bf{w}},
\end{align}
where $\mathbf{w}=[w_1,w_2]^T$. From (\ref{y}), the optimal ML detector can be formulated as
\begin{align}\label{ML}
\left( {\hat l,{{\hat \alpha }_1},{{\hat \alpha }_2}} \right) = \mathop {\arg \min }\limits_{l,{\alpha _1},{\alpha _2}} {\left\| {{\bf{y}} - {\bf{C}}{{\bf{h}}_l}} \right\|^2},
\end{align}
where $\hat{l}$, $\hat{\alpha}_1$, and $\hat{\alpha}_2$ are the estimates of $l$, $\alpha_1$, and $\alpha_2$, respectively.
Unfortunately, the computational complexity of the optimal ML detector in (\ref{ML}) in terms of complex multiplications is $\mathcal{O}(N_tM^2)$, which becomes impractical for large values of $N_t$ and $M$. The design of low-complexity ML detector will be investigated as follows.

Given a realization of $l\in\{1,\ldots,N_t\}$, according to the classical Alamouti principle, we can obtain the combined signals as follows:
\begin{align}\label{r_12}
({r_1})_l &= {y_1}h_{1l}^* + y_2^*{h_{2l}} \nonumber \\
 &= \left( {{{\left| {{h_{1l}}} \right|}^2} + {{\left| {{h_{2l}}} \right|}^2}} \right)\exp (j{\alpha _1}) + h_{1l}^*{w_1} + {h_{2l}}w_2^*, \\
({r_2})_l &= {y_1}h_{2l}^* - y_2^*{h_{1l}} \nonumber \\
 &= \left( {{{\left| {{h_{1l}}} \right|}^2} + {{\left| {{h_{2l}}} \right|}^2}} \right)\exp (j{\alpha _2}) - h_{1l}{w_2^*} + {h_{2l}^*}w_1.
\end{align}
Then, in the case of the $l$-th antenna being activated, the ML metric can be calculated by
\begin{align}\label{D}
D(l) &= \mathop {\min }\limits_{{\alpha _1} \in \mathcal{M}} {\left| {({r_1})_l - \left( {{{\left| {{h_{1l}}} \right|}^2} + {{\left| {{h_{2l}}} \right|}^2}} \right)\exp (j{\alpha _1})} \right|^2} \nonumber \\
& \quad + \mathop {\min }\limits_{{\alpha _2} \in \mathcal{M}} {\left| {({r_2})_l - \left( {{{\left| {{h_{1l}}} \right|}^2} + {{\left| {{h_{2l}}} \right|}^2}} \right)\exp (j{\alpha _2})} \right|^2}.
\end{align}
Finally, the ML receiver makes a decision onto the active antenna index from
\begin{align}
\hat l = \mathop {\arg \min }\limits_l D(l),
\end{align}
and recovers $\alpha_1$ and $\alpha_2$ from
\begin{align}\label{hat_1}
{{\hat \alpha }_1} = \mathop {\arg \min }\limits_{{\alpha _1} \in \mathcal{M}} {\left| {({r_1})_{\hat{l}} - \left( {{{\left| {{h_{1\hat{l}}}} \right|}^2} + {{\left| {{h_{2\hat{l}}}} \right|}^2}} \right)\exp (j{\alpha _1})} \right|^2},
\end{align}
\begin{align}\label{hat_2}
{{\hat \alpha }_2}= \mathop {\arg \min }\limits_{{\alpha _2} \in \mathcal{M}} {\left| {({r_2})_{\hat{l}} - \left( {{{\left| {{h_{1\hat{l}}}} \right|}^2} + {{\left| {{h_{2\hat{l}}}} \right|}^2}} \right)\exp (j{\alpha _2})} \right|^2}.
\end{align}
As seen from (\ref{D}), the computational complexity of this detector in terms of complex multiplications is reduced to $\mathcal{O}(2N_tM)$.

\subsection{Performance Analysis}
In this subsection, approximate and asymptotic ABEP expressions are derived in closed-form for the source and the RIS of RIS-SSK-ASTBC utilizing the optimal ML detector.

\subsubsection{ABEP of the Source}
The source information is completely conveyed through the active antenna index. Let us first study the conditional PEP from (\ref{ML}), which is the probability of detecting $(l,\mathbf{C})$ as $(\hat{l},\hat{\mathbf{C}})$ conditioned on $\mathbf{G}$ and $\mathbf{f}$, namely
\begin{align}\label{CPEP}
&\Pr\left( {\left( {l,{\bf{C}}} \right) \to ( {\hat l,\hat{\bf{C}}} )\left| {{\bf{G}},{\bf{f}}} \right.} \right) \nonumber \\
&= \Pr \left( {{{\left\| {{\bf{y}} - {\bf{C}}{{\bf{h}}_l}} \right\|}^2} > {{\left\| {{\bf{y}} - \hat{\bf{C}}{{\bf{h}}_{\hat l}}} \right\|}^2}} \right) \nonumber \\
& = Q\left( {\sqrt {\frac{{{\rho{\left\| {{\bf{C}}{{\bf{h}}_l} - \hat{\bf{ C}}{{\bf{h}}_{\hat l}}} \right\|}^2}}}{{2}}} } \right).
\end{align}
Here, we resort to the CLT under the assumption of $N \gg 1$ for the calculation of this PEP. Under the CLT, each entity of $\mathbf{Ch}_l$ and $\hat{\mathbf{C}}\mathbf{h}_{\hat{l}}$ can be treated as a random variable following the distribution $\mathcal{CN}(0,N)$. Hence, when $l \neq \hat{l}$, it can be shown that $\| {{\bf{C}}{{\bf{h}}_l} - \hat{\bf{C}}{{\bf{h}}_{\hat l}}}\|^2$ is a central chi-square random variable with four degrees of freedom, whose PDF is given by
\begin{align}\label{PDF}
f\left( x \right) = \frac{x}{{A{N^2}}}\exp \left( { - \frac{x}{{BN}}} \right),
\end{align}
where $A=4$ and $B=2$. Averaging $\Pr( {( {l,{\bf{C}}}) \to ( {\hat l,\hat{\bf{C}}} )| {{\bf{G}},{\bf{f}}}}) $ over $\|{\bf{C}}{{\bf{h}}_l} - \hat{\bf{C}}{{\bf{h}}_{\hat l}}\|^2$ results in the following unconditional PEP [\ref{Alouini}, Eq. (64)]:
\begin{align}\label{UPEP}
\Pr \left( {\left( {l,{\bf{C}}} \right) \to ( {\hat l,\hat{\bf{C}}})} \right) &= \int_0^{ + \infty } {Q\left( {\sqrt {\frac{\rho x}{{2}}} } \right)} f\left( x \right)dx \nonumber \\
 &= 3{p^2} - 2{p^3},
\end{align}
where
\begin{align}
p=\frac{1}{2}\left( {1 - \sqrt {\frac{N\rho}{{2 + N\rho}}} } \right).
\end{align}
According to the union bounding technique, an upper bound on the error probability of active antenna index detection is given by
\begin{align}\label{P_I}
{P_e} &\le \frac{1}{{{N_t}{M^2}}}\sum\limits_{\scriptstyle l,\hat l\hfill\atop
\scriptstyle l \ne \hat l\hfill} {\sum\limits_{{\bf{C}},{\bf{\hat C}}} {\Pr \left( {\left( {l,{\bf{C}}} \right) \to ( {\hat l,{\bf{\hat C}}})} \right)} } \nonumber \\
&= {M^2}\left( {{N_t} - 1} \right)\left( {3{p^2} - 2{p^3}} \right).
\end{align}
Finally, the ABEP of the source can be expressed as [\ref{Li}, Eq. (13)]
\begin{align}\label{P_bS}
{P_{bS}} \approx \frac{1}{2}{P_e} \cdot \frac{{{N_t}}}{{{N_t} - 1}}.
\end{align}

\subsubsection{ABEP of the RIS}
It is obvious that the error events for phase shift detection can be categorized into two complementary types, depending on whether the index of the active antenna is detected correctly or not. Thus, the overall ABEP of the RIS can be given by
\begin{align}\label{P_bRIS}
{P_{bRIS}} \approx \frac{1}{2}{P_e} + (1 - {P_e}){P_A},
\end{align}
where $P_e$ is given in (\ref{P_I}), $P_A$ is the ABEP of $M$-PSK demodulation in the case of correct detection for the active antenna index, and the factor $1/2$ accounts for the ABEP in the case of incorrect detection for the active antenna index.

For the calculation of $P_A$, the instantaneous receive SNR per virtual PSK symbol can be derived from (\ref{hat_1}) or (\ref{hat_2}) as $\rho {\left\| {{{\bf{h}}_l}} \right\|^2}$. Therefore, the conditional ABEP of $M$-PSK demodulation can be expressed as [\ref{Simon}, Eq. (8.23)]
\begin{align}
{P_{A\left| {{{\left\| {{{\bf{h}}_l}} \right\|}^2}} \right.}} & \cong \frac{2}{{\max \left( {{{\log }_2}\left( M \right),2} \right)}} \nonumber \\
&\quad \times \sum\limits_{i = 1}^{\max \left( {M/4,1} \right)} {Q\left( {\sqrt {2\rho g_{\textrm{PSK}}(i) {{\left\| {{{\bf{h}}_l}} \right\|}^2}} } \right)},
\end{align}
where $g_{\textrm{PSK}}(i)={\sin}^2((2i - 1)\pi/M)$. By resorting to the CLT, $||\mathbf{h}_l||^2$ can be approximated as a central chi-square random variable with four degrees of freedom, whose PDF is given by (\ref{PDF}) with $A=1/4$ and $B=1/2$.
Then, by averaging ${P_{A\left| {{{\left\| {{{\bf{h}}_l}} \right\|}^2}} \right.}}$ over $||\mathbf{h}_l||^2$, we have
\begin{align}\label{P_A}
{P_A} \cong \frac{2}{{\max \left( {{{\log }_2}\left( M \right),2} \right)}} \times \sum\limits_{i = 1}^{\max \left( {M/4,1} \right)} {3{{\left[ {q(i)} \right]}^2} - 2{{\left[ {q(i)} \right]}^3}},
\end{align}
where
\begin{align}
q(i) = \frac{1}{2}\left(1 - \sqrt {\frac{{\rho g_{\textrm{PSK}}(i) N}}{{2 + \rho {g_{\textrm{PSK}}(i) N}}} } \right).
\end{align}
Finally, by substituting (\ref{P_I}) and (\ref{P_A}) into (\ref{P_bRIS}), we obtain the ABEP of the RIS.

\subsubsection{Asymptotic Analysis}
By taking the Taylor series of the exponential function and ignoring higher order terms, (\ref{PDF}) reduces to
\begin{align}\label{Simp_PDF}
f\left( x \right) \approx \frac{x}{{4{N^2}}}.
\end{align}
With the simplified PDF in (\ref{Simp_PDF}), the unconditional PEP in (\ref{UPEP}) can be re-calculated asymptotically as
\begin{align}\label{Asy_PEP}
\Pr \left( {\left( {l,{\bf{C}}} \right) \to ( {\hat l,\hat{\bf{ C}}} )} \right) \to \frac{3}{4} \cdot {\left( {\rho N} \right)^{ - 2}}.
\end{align}
Putting (\ref{Asy_PEP}) into (\ref{P_I}) yields
\begin{align}\label{Asy_P_e}
{P_e} \le \frac{3}{4}{M^2}\left( {{N_t} - 1} \right){\left( {\rho N} \right)^{ - 2}}.
\end{align}
Then, by substituting (\ref{Asy_P_e}) into (\ref{P_bS}), we have the asymptotic ABEP of the source as follows:
\begin{align}\label{Asy_P_bS}
{P_{bS}} \approx \frac{3}{8}{M^2}{N_t}{\left( {\rho N} \right)^{ - 2}},
\end{align}
which achieves a diversity order of two. Interestingly, we observe from (\ref{Asy_P_bS}) that the achievable diversity order of the source is irrelevant to $N$, namely the number of reflecting elements. However, increasing $N$ is equivalent to increasing $\rho$, both of which decrease the PEP quadratically. These observations also apply to the RIS. Similar to (\ref{Asy_PEP}), we have
\begin{align}\label{Asy_P_A}
{P_A} \to \frac{3}{{2\cdot \max \left( {{{\log }_2}\left( M \right),2} \right)}} \times \sum\limits_{i = 1}^{\max \left( {M/4,1} \right)} \left( {\rho g_{\textrm{PSK}}(i) N} \right)^{ - 2}.
\end{align}
Further, at high SNR, (\ref{P_bRIS}) can be simplified to
\begin{align}\label{Asy_P_bRIS}
{P_{bRIS}} \approx \frac{1}{2}{P_e} + {P_A}.
\end{align}
Substituting (\ref{Asy_P_e}) and (\ref{Asy_P_A}) into (\ref{Asy_P_bRIS}) results in the asymptotic ABEP of the source. Obviously, we observe from $P_{bS}$ and $P_{bRIS}$ that the information bits from both the source and the RIS have two-diversity-order protection with increasing SNR or the number of reflecting elements. Moreover, doubling $N$ is equivalent to doubling $\rho$, such that a consistent SNR gain of about 3 dB gain can be achieved every time $N$ is doubled for both the source and the RIS.

\textit{Remark 2:} In RIS-SSK-ASTBC, the reflection coefficients completely depend on the information to be transmitted from the RIS, and no beamforming is performed at the RIS. Hence, the CSI is not needed at the RIS. Actually, if the CSI is available at the RIS, we can implement the Alamouti transmission and beamforming simultaneously at the RIS, further enhancing the performance of RIS-SSK-ASTBC.

\section{Simulation Results and Comparisons}
In this section, we perform Monte Carlo simulations to assess the uncoded BER performance of RIS-SSK-PB and RIS-SSK-ASTBC. In all simulations, we plot
the BER versus $\rho = 1/N_0$. The channels are assumed to be Rayleigh flat fading channels, the CSI is perfectly known to the destination and/or the RIS, and any path loss effect is neglected. For RIS-SSK-PB, the traditional SSK \cite{Jeganathan2009Space} and the intelligent RIS-SSK \cite{Canbilen2020Reconfigurable} are taken as reference schemes. The amplify-and-forward aided SSK and blind RIS-SSK are excluded from the performance comparison, since they have been shown to be inferior to the intelligent RIS-SSK in \cite{Canbilen2020Reconfigurable}. For RIS-SSK-ASTBC, RIS-SSK-VBLAST that employs the two-antenna based VBLAST scheme to transmit RIS's private information is chosen as a benchmark. All considered schemes employ single-antenna ML detection for fair comparisons. Each BER point is obtained by averaging over at least $10^5$ transmission.

\subsection{RIS-SSK-PB}
\begin{figure}[!t]
	\centering
	\includegraphics[width=3.7in]{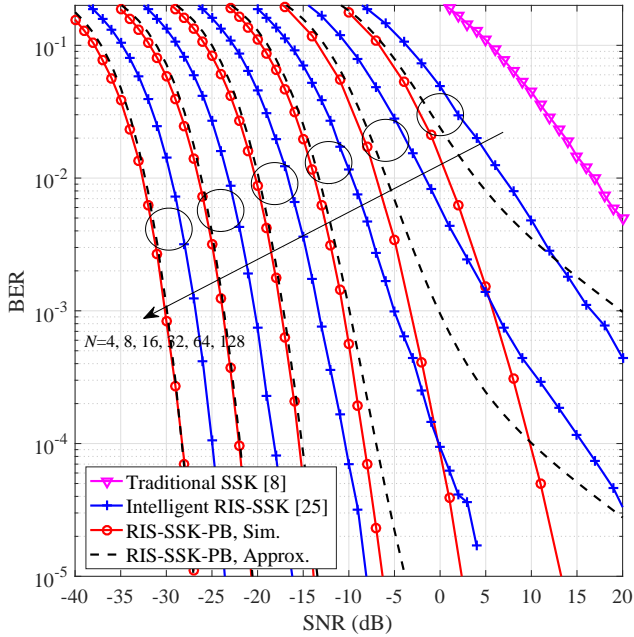}
	\caption{Performance comparison among traditional SSK, intelligent RIS-SSK, and RIS-SSK-PB, where $N_t=2$ and $N=4,8,16,32,64,128$.}
	\label{Fig3}
\end{figure}
Fig.~\ref{Fig3} depicts the comparison results among traditional SSK, intelligent RIS-SSK, and RIS-SSK-PB, where $N_t=2$ and $N=4,8,16,32,64,128$. The approximate ABEP curves derived in Section II.C are also presented for RIS-SSK-PB. As shown in Fig.~\ref{Fig3}, due to the nature of the CLT, the analytical results become more accurate with increasing $N$. When $N \ge 32$, the analytical ABEP curves can predict the simulated counterparts very well. With increasing $N$, the BER performance of both intelligent RIS-SSK and RIS-SSK-PB improve, and both perform much better than traditional SSK for all $N$ in the overall SNR region. Moreover, RIS-SSK-PB significantly outperforms the intelligent RIS-SSK for all $N$ and SNR values. In particular, RIS-SSK-PB achieves an additional diversity gain over the intelligent RIS-SSK, and the gain becomes more prominent when $N$ is a smaller value. Fortunately, even for $N=128$, approximately 3 dB SNR gain is obtained by RIS-SSK-PB over the intelligent RIS-SSK, at a BER value of $10^{-4}$. Note that these advantages are achieved along with the benefits of avoiding the requirements of active antenna index information and online phase adjustments.

\begin{figure}[!t]
	\centering
	\includegraphics[width=3.7in]{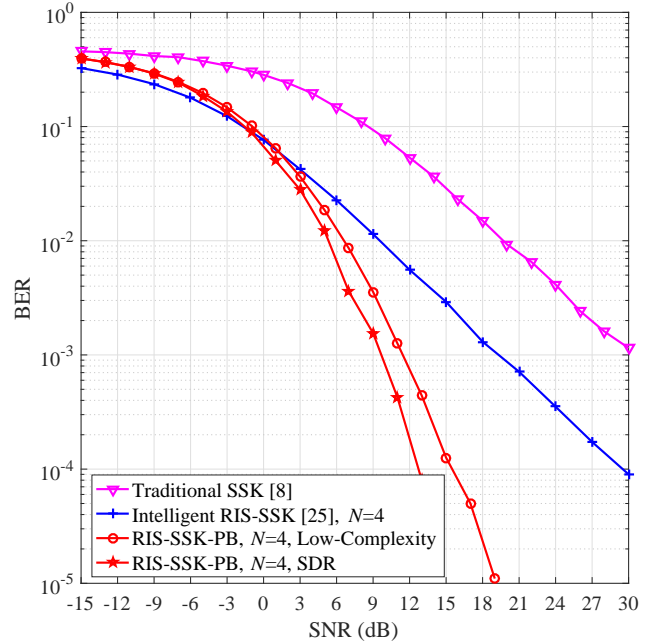}
	\caption{Performance comparison among traditional SSK, intelligent RIS-SSK, and RIS-SSK-PB, where $N_t=4$ and $N=4$.}
	\label{Fig4}
\end{figure}

In Fig.~\ref{Fig4}, we make comparisons among traditional SSK, intelligent RIS-SSK, and RIS-SSK-PB, where $N_t=4$ and $N=4$. For RIS-SSK-PB, both the SDR and low-complexity beamforming algorithms are taken into account. The random simulation times in the SDR method for solving problem (P3) is set as 100. As seen from Fig.~\ref{Fig4}, with the aid of a 4-element RIS, both the intelligent RIS-SSK and RIS-SSK-PB perform much better than traditional SSK throughout the SNR region. Moreover, no matter which beamforming algorithm is employed, RIS-SSK-PB achieves a higher diversity gain than the intelligent RIS-SSK and traditional SSK. In particular, RIS-SSK-PB with the SDR method outperforms the intelligent RIS-SSK when SNR is greater than -1 dB, achieving approximately 19 dB SNR gain at a BER value of $10^{-4}$. Notably, the low-complexity beamformer suffers from about 3 dB performance loss as compared with the SDR method at a BER value of $10^{-4}$. However, RIS-SSK-PB with the low-complexity beamforming is still superior to the intelligent RIS-SSK and traditional SSK with much reduced complexity.

\subsection{RIS-SSK-ASTBC}
\begin{figure}[!t]
	\centering
	\includegraphics[width=3.5in]{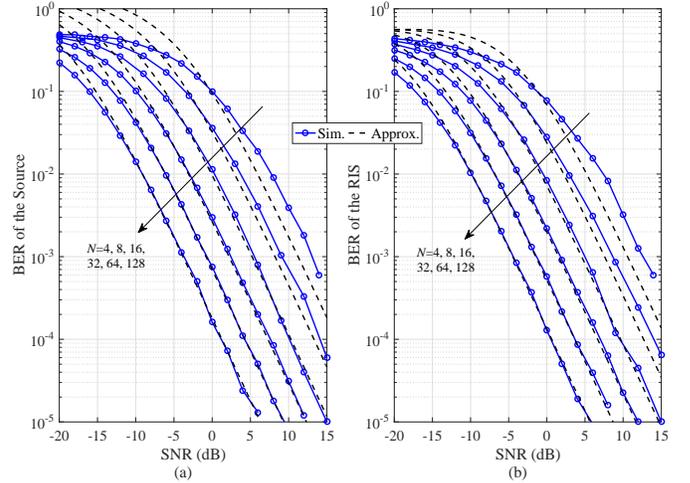}
	\caption{Performance of RIS-SSK-ASTBC for: (a) the source and (b) the RIS, where $N_t=2$, $M=2$, and $N=4,8,16,32,64,128$.}
	\label{Fig5}
\end{figure}

Fig.~\ref{Fig5} presents the BER performance of RIS-SSK-ASTBC, where $N_t=2$, $M=2$, and $N=4,8,16,32,64,128$. To verify the analysis given in Section III.B, we also plot the approximate ABEP curves for the source, namely (\ref{P_bS}), and for the RIS, namely (\ref{P_bRIS}). As seen from Fig.~\ref{Fig5}, since we resort to the CLT in the performance analysis, the analytical results become accurate with increasing $N$ for both the source and the RIS. In the cases of $N \ge 32$, the theoretical curves agree with the simulated counterparts very well in the SNR region of interest. It can be seen that increasing $N$ yields performance improvement for both the source and the RIS. For example, about 3 dB SNR gain is observed at a BER value of $10^{-4}$ for the source and the RIS, when $N$ increases from 64 to 128. However, there is no diversity improvement, since the source and the RIS achieve a diversity order of two, which is irrelevant to the value of $N$.

\begin{figure}[!t]
	\centering
	\includegraphics[width=3.5in]{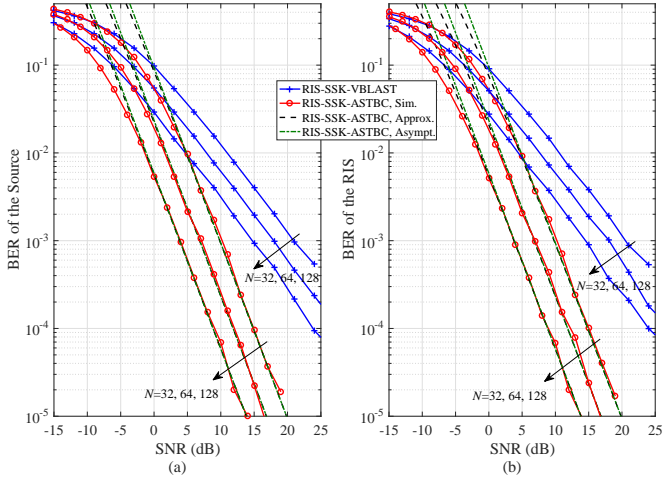}
	\caption{Performance comparison between RIS-SSK-ASTBC with $M=8$ and RIS-SSK-VBLAST with $M=2$ for: (a) the source and (b) the RIS, where $N_t=4$ and $N=32,64,128$.}
	\label{Fig6}
\end{figure}

Fig.~\ref{Fig6} shows the performance comparison between RIS-SSK-ASTBC with $M=8$, and RIS-SSK-VBLAST with $M=2$, where $N_t=4$ and $N=32,64,128$. All considered schemes achieve the same spectral efficiency of 4 bpcu. Both the approximate and asymptotic results are also provided. From Fig.~\ref{Fig6}, we observe that the approximate and asymptotic ABEP curves match the simulated counterparts very well. For both the source and the RIS, RIS-SSK-ASTBC achieves a diversity order of two, while RIS-SSK-VBALST achieves a diversity order of unity. Due to the higher diversity order, RIS-SSK-ASTBC significantly outperforms RIS-SSK-VBLAST, though RIS-SSK-ASTBC employs a higher constellation size for achieving the same spectral efficiency as RIS-SSK-VBLAST. For instance, the source and the RIS of RIS-SSK-ASTBC achieve about 15 dB SNR gain over those counterparts of RIS-SSK-VBLAST, at a BER value of $10^{-4}$.

\section{Conclusion}
In this paper, we have proposed RIS-SSK-PB and RIS-SSK-ASTBC schemes to improve the error performance and throughput of traditional SSK systems, respectively. In RIS-SSK-PB, the RIS is employed for beamforming, which maximizes the minimum squared Euclidean distance between any two decision points. The SDR and low-complexity algorithms have been developed for the reflecting optimization. In RIS-SSK-ASTBC, the RIS is employed for ASTBC, which enables the RIS to transmits its own Alamouti-coded information while reflecting the incident signals. A low-complexity ML detector has been studied for RIS-SSK-ASTBC. The approximate ABEP expressions have been derived for the source and/or the RIS in closed-form assuming ML detection. Computer simulations have corroborated the performance analysis and the performance improvement achieved by RIS-SSK-PB and RIS-SSK-ASTBC. It could be concluded that the two proposed schemes are viable candidates for energy-efficient and ultra-reliable communications. Also, RIS-SSK-ASTBC can be extended by employing general STBC with more than two sub-surfaces at the RIS. For future work, we intend to extend the proposed RIS-SSK schemes to more practical RIS models, including the RIS unit cell model of \cite{Abeywickrama2020Intelligent} and the mutual-impedances-based end-to-end model of \cite{Gradoni2020End}.


\begin{thebibliography}{99}
\bibitem{Wen2017Index}
M. Wen, X. Cheng, and L. Yang, \textit{Index Modulation for 5G Wireless Communications}. Springer International Publishing AG, Cham, Switzerland, 2017.

\bibitem{Cheng2018Index}
X. Cheng, M. Zhang, M. Wen, and L. Yang, ``Index modulation for 5G: Striving to do more with less,'' \textit{IEEE Wireless Commun. Mag.}, vol. 25, no. 2, pp. 126--132, Apr.~2018.

\bibitem{Basar2017Index}
E. Basar, M. Wen, R. Mesleh, M. Di Renzo, Y. Xiao, and H. Haas, ``Index modulation techniques for next-generation wireless networks,'' \textit{IEEE Access}, vol.~5, pp.~16693--16746, Sep.~2017.

\bibitem{Li2020Subcarrier}
Q. Li, M. Wen, B. Clerckx, S. Mumtaz, A. Al-Dulaimi, and R. Q. Hu, ``Subcarrier index modulation for future wireless networks: Principles, applications, and challenges,'' \textit{IEEE Wireless Commun.}, vol.~27, no.~3, pp.~118--125, Jun.~ 2020.

\bibitem{Wen2019A}
M. Wen, B. Zheng, K. J. Kim, M. Di Renzo, T. A. Tsiftsis, K. Chen, and N. Al-Dhahir, ``A survey on spatial modulation in emerging wireless systems: Research progresses and applications,'' \textit{IEEE J. Select. Areas Commun.}, vol.~37, no.~9, pp. 1949--1972, Sep. 2019.

\bibitem{Renzo2014Spatial}
M. Di Renzo, H. Haas, A. Ghrayeb, S. Sugiura, and L. Hanzo, ``Spatial modulation for generalized MIMO: Challenges, opportunities and implementation,'' \textit{Proc. IEEE}, vol. 102, no. 1, pp. 56--103, Jan.~2014.


\bibitem{Mesleh2008Spatial}
R. Y. Mesleh, H. Haas, S. Sinanovic, C. W. Ahn, and S. Yun, ``Spatial modulation,'' \textit{IEEE Trans. Veh. Technol.}, vol.~57, no.~4, pp.~2228--2241,
Jul.~2008.


\bibitem{Jeganathan2009Space}
J. Jeganathan, A. Ghrayeb, L. Szczecinski, and A. Ceron, ``Space shift keying modulation for MIMO channels,'' \textit{IEEE Trans. Wireless Commun.}, vol. 8, no. 7, pp. 3692--3703, Jul. 2009.


\bibitem{Renzo2020Smart}
M. Di Renzo, A. Zappone, M. Debbah, M.-S. Alouini, C. Yuen, J. De Rosny, and S. Tretyakov, ``Smart radio environments empowered by reconfigurable intelligent surfaces: How it works, state of research, and road ahead,'' \textit{IEEE J. Sel. Areas Commun.}, vol. 38, no. 11, pp.~2450--2525, Nov. 2020.


\bibitem{Renzo2019Smart}
M. Di Renzo, M. Debbah, D.-T. Phan-Huy, A. Zappone, M.-S. Alouini, C. Yuen, V. Sciancalepore, G. C. Alexandropoulos, J. Hoydis, H. Gacanin, J. de Rosny, A. Bounceur, G. Lerosey, and M. Fink, ``Smart radio environments empowered by reconfigurable AI meta-surfaces: An idea whose time has come,'' \textit{EURASIP J. Wireless Commun. Netw.}, vol.~2019, no.~1, pp. 1--20, May 2019.

\bibitem{Basar2019Wireless}
E. Basar, M. Di Renzo, J. De Rosny, M. Debbah, M. Alouini, and R. Zhang, ``Wireless communications through reconfigurable intelligent surfaces,'' \textit{IEEE Access}, vol. 7, pp. 116753--116773, 2019.

\bibitem{Huang2020Holographic}
C. Huang, S. Hu, G. C. Alexandropoulos, A. Zappone, C. Yuen, R. Zhang, M. Di Renzo, and M. Debbah, ``Holographic MIMO surfaces for 6G wireless networks: Opportunities, challenges, and trends,'' \textit{IEEE Wireless Communications}, vol. 27, no. 5, pp. 118--125, Oct. 2020.


\bibitem{Tang2020Wireless}
W. Tang, M. Z. Chen, J. Y. Dai, Y. Zeng, X. Zhao, S. Jin, Q. Cheng, and T. J. Cui, ``Wireless communications with programmable metasurface: New paradigms, opportunities, and challenges on transceiver design,'' \textit{IEEE Wireless Commun.}, vol. 27, no. 2, pp.~180--187, Apr. 2020.

\bibitem{Basar2019Transmission}
E. Basar, ``Transmission through large intelligent surfaces: A new frontier in wireless communications,'' in \textit{Proc. European Conf. Netw. Commun.
(EuCNC)}, Valencia, Spain, Jun. 2019, pp. 112--117.


\bibitem{Ye2020Joint}
J. Ye, S. Guo, and M.-S. Alouini, ``Joint reflecting and precoding designs for SER minimization in reconfigurable intelligent surfaces assisted MIMO systems,'' \textit{IEEE Trans. Wireless Commun.}, vol. 19, no.~8, pp.~5561--5574, Aug. 2020.


\bibitem{Yigit2020Low}
Z. Yigit, E. Basar, and I. Altunbas, ``Low complexity adaptation for reconfigurable intelligent surface-based MIMO systems,'' \textit{IEEE Commun. Lett.}, vol. 24, no. 12, pp. 2946--2950, Dec. 2020.

\bibitem{Dong2020Enhancing}
L. Dong and H. Wang, ``Enhancing secure MIMO transmission via intelligent reflecting surface,'' \textit{IEEE Trans. Wireless Commun.},  vol. 19, no. 11, pp. 7543--7556, Nov. 2020.


\bibitem{Wu2019Intelligent}
Q. Wu and R. Zhang, ``Intelligent reflecting surface enhanced wireless network via joint active and passive beamforming,'' \textit{IEEE Trans. Wireless Commun.}, vol. 18, no. 11, pp. 5394--5409, Nov. 2019.

\bibitem{Wu2020Beamforming}
Q. Wu and R. Zhang, ``Beamforming optimization for wireless network aided by intelligent reflecting surface with discrete phase shifts,'' \textit{IEEE
Trans. Commun.}, vol.~68, no.~3, pp.~1838--1851, Mar.~2020.


\bibitem{Huang2019Reconfigurable}
C. Huang, A. Zappone, G. C. Alexandropoulos, M. Debbah, and C. Yuen, ``Reconfigurable intelligent surfaces for energy efficiency in wireless communication,'' \textit{IEEE Trans. Wireless Commun.}, vol.~18, no.~8, pp.~4157--4170, Aug.~2019.

\bibitem{Huang2018Energy}
C. Huang, G. C. Alexandropoulos, A. Zappone, M. Debbah, and C. Yuen, ``Energy efficient multi-user MISO communication using low resolution large intelligent surfaces,'' in \textit{Proc. Globecom Workshops}, Abu Dhabi, United Arab Emirates, Dec. 2018, pp.~1--6.

\bibitem{Hou2020Reconfigurable}
T. Hou, Y. Liu, Z. Song, X. Sun, Y. Chen, and L. Hanzo, ``Reconfigurable intelligent surface aided NOMA networks,'' \textit{IEEE J. Sel. Areas Commun.}, vol. 38, no. 11, pp. 2575--2588, Nov. 2020.


\bibitem{Zheng2020Intelligent}
B. Zheng and R. Zhang, ``Intelligent reflecting surface-enhanced OFDM: Channel estimation and reflection optimization,'' \textit{IEEE Wireless Commun. Lett.}, vol. 9, no. 4, pp. 518--522, Apr. 2020.

\bibitem{Bouida2020Reconfigurable}
Z. Bouida, H. El-Sallabi, A. Ghrayeb, and K. A. Qaraqe, ``Reconfigurable antenna-based space-shift keying (SSK) for MIMO Rician channels,'' \textit{IEEE Trans. Wireless Commun}, vol.~15, no.~1, pp.~446--457, Jan.~2016.

\bibitem{Canbilen2020Reconfigurable}
A. E. Canbilen, E. Basar, and S. S. Ikki, ``Reconfigurable intelligent surface-assisted space shift keying,'' \textit{IEEE Wireless Commun. Lett.}, vol.~9, no. 9, pp. 1495--1499, Sep.~2020.

\bibitem{Li2020Single}
Q. Li, M. Wen, and M. D. Renzo, ``Single-RF MIMO: From spatial modulation to metasurface-based modulation'', Available: https://arxiv.org/abs/2009.00789, accessed on Sep. 2020.

\bibitem{Karasik2020Beyond}
R. Karasik, O. Simeone, M. Di Renzo, and S. Shamai Shitz, ``Beyond max-SNR: Joint encoding for reconfigurable intelligent surfaces,'' \textit{2020 IEEE International Symposium on Information Theory (ISIT)}, Los Angeles, CA, USA, 2020, pp. 2965--2970.


\bibitem{Tang2020MIMO}
W. Tang, J. Y. Dai, M. Z. Chen, K.-K. Wong, X. Li, X. Zhao, S. Jin, Q. Cheng, and T. J. Cui, ``MIMO transmission through reconfigurable intelligent surface: System design, analysis, and implementation,'' \textit{IEEE J. Sel. Areas Commun.}, vol. 38, no. 11, pp. 2683--2699, Nov. 2020.


\bibitem{Guo2020Reflecting}
S. Guo, S. Lv, H. Zhang, J. Ye, and P. Zhang, ``Reflecting modulation,'' \textit{IEEE J. Sel. Areas Commun.}, vol. 38, no. 11, pp. 2548--2561, Nov. 2020.


\bibitem{Yan2020Passive}
W. Yan, X. Yuan, Z. He, and X. Kuai, ``Passive beamforming and information transfer design for reconfigurable intelligent surfaces aided multiuser MIMO systems,'' \textit{IEEE J. Sel. Areas Commun.}, vol. 38, no. 8, pp. 1793--1808, Aug. 2020.

\bibitem{Lin2020Reconfigurable}
S. Lin, B. Zheng, G. C. Alexandropoulos, M. Wen, M. Di Renzo, and F. Chen, ``Reconfigurable intelligent surfaces with reflection pattern modulation: Beamforming design and performance analysis,'' \textit{IEEE Trans. Wireless Commun.}, to be published, 2020.


\bibitem{Basar2020Reconfigurable}
E. Basar, ``Reconfigurable intelligent surface-based index modulation: A new beyond MIMO paradigm for 6G,'' \textit{IEEE Trans. Commun.}, vol.~68, no.~5, pp.~3187--3196, May 2020.

\bibitem{Ma2020Large}
T. Ma, Y. Xiao, X. Lei, P. Yang, X. Lei, and O. A. Dobre, ``Large intelligent surface assisted wireless communications with spatial modulation and antenna selection,'' \textit{IEEE J. Sel. Areas Commun.}, vol. 38, no. 11, pp. 2562--2574, Nov.~2020.

\bibitem{Khaleel2020Reconfigurable}
A. Khaleel and E. Basar, ``Reconfigurable intelligent surface-empowered MIMO systems,'' \textit{IEEE Systems Journal}, to be published, 2020.

\bibitem{Bjornson2020Rayleigh}
E. Bjornson and L. Sanguinetti, ``Rayleigh fading modeling and channel hardening for reconfigurable intelligent surfaces,'' \textit{IEEE Wireless Commun. Lett.}, to be published, 2021.

\bibitem{Alexandropoulos2020A}
G. C. Alexandropoulos and E. Vlachos, ``A hardware architecture for reconfigurable intelligent surfaces with minimal active elements for explicit channel estimation,'' in \textit{Proc. IEEE ICASSP}, Barcelona, Spain, May 2020, pp. 9175-9179.

\bibitem{Proakis2005Digital}
J. G. Proakis and M. Salehi, \textit{Digital Communications}, 5th ed. New York, NY, USA: McGraw-Hill, 2008.

\bibitem{Grant2016Intelligent}
M. Grant and S. Boyd, \textit{CVX: MATLAB software for disciplined convex programming}. 2016. [Online]. Available: http://cvxr.com/cvx.

\bibitem{Boulogeorgos2020Performance}
A.-A. A. Boulogeorgos and A. Alexiou, ``Performance analysis of reconfigurable intelligent surface-assisted wireless systems and comparison with relaying,'' \textit{IEEE Access}, vol. 8, pp. 94463--94483, 2020.

\bibitem{Chiani2003New}
M. Chiani, D. Dardari, and M. K. Simon, ``New exponential bounds and approximations for the computation of error probability in fading channels,'' \textit{IEEE Trans. Wireless Commun.}, vol. 2, no. 4, pp. 840--845, Jul. 2003.


\bibitem{Alouini1999}\label{Alouini}
M.-S. Alouini and A. J. Goldsmith, ``A unified approach for calculating error rates of linearly modulated signals over generalized fading channels,'' \textit{IEEE Trans. Commun.}, vol.~47, no.~9, pp.~1324--1334, Sep.~1999.

\bibitem{Li2020Dual}\label{Li}
Q. Li, M. Wen, M. D. Renzo, H. V. Poor, S. Mumtaz, and F. Chen, ``Dual-hop spatial modulation with a relay transmitting its own information,'' \textit{IEEE Trans. Wireless Commun.}, vol.~19, no.~7, pp.~4449--4463, Jul. 2020.

\bibitem{Simon2005Digital}\label{Simon}
M. K. Simon and M.-S. Alouini, \textit{Digital Communication Over Fading Channels}, 2nd ed. New York, NY, USA: Wiley, 2005.


\bibitem{Abeywickrama2020Intelligent}
S. Abeywickrama, R. Zhang, Q. Wu, and C. Yuen, ``Intelligent reflecting surface: Practical phase shift model and beamforming optimization,'' \textit{IEEE Trans. Commun.}, vol. 68, no. 9, pp. 5849--5863, Sep. 2020.

\bibitem{Gradoni2020End}
G. Gradoni and M. Di Renzo, ``End-to-end mutual coupling aware communication model for reconfigurable intelligent surfaces: An electromagnetic-compliant approach based on mutual impedances,'' Available: https://arxiv.org/abs/2009.02694, accessed on Sep. 2020.


%\bibitem{Alamouti1998A}
%S. Alamouti, ``A simple transmit diversity technique for wireless communications,'' \textit{IEEE J. Sel. Areas Commun.}, vol.~16, no.~8, pp.~1451--1458,
%Oct.~1998.
\end{thebibliography}
\end{document}